\documentclass[12pt]{iopart}
\usepackage{times} 

\begin{document}

\def\SLC{{\Omega h^2}_{\rm SLC}}
\def\SQC{{\Omega h^2}_{\rm SQC}}
\def\NCC{{\Omega h^2}_{\rm NCC}}
\def\ACC{{\Omega h^2}_{\rm ACC}}
\def\IGC{{\Omega h^2}_{\rm IGC}}
\def\COA{{\Omega h^2}_{\rm COA}} 
\def\mchi{m_{\chi} }
\def\gev{\mbox{ GeV}} 
\def\tev{\mbox{ TeV}}
\def\dtgev{\,\mbox{GeV}^{-2}}
\title{Comparison of coannihilation effects in low-energy MSSM}

\author{V.A.~Bednyakov}
\address{Laboratory of Nuclear Problems,
         Joint Institute for Nuclear Research, \\ 
         141980 Dubna, Russia; E-mail: bedny@nusun.jinr.ru}

\begin{abstract} 
	The neutralino relic density is calculated
	in the low-energy effective 
	Minimal Supersymmetric extension of Standard Model (effMSSM). 
	The slepton-neutralino,
	squark-neutralino and neutralino/chargino-neutralino 
	coannihilation channels are taking into account. 
	The comparative study of these 
	coannihilations is performed and 
	demonstrated that all of them give 
	the sizable contributions to the reduction of the 
	neutralino relic density.
	It is shown that the predictions for 
	direct dark matter detection rates 
	are not strongly affected by
	these coannihilation channels in the effMSSM. 
\end{abstract} 

\section{Introduction} 
	A variety of data ranging from galactic rotation curves to 
	large scale structure formation and the cosmic microwave background
	radiation imply a significant density 
	$0.1< \Omega h^2 <0.3$  
\cite{acc} of so-called cold dark matter (CDM). 
	Here $\Omega=\rho/\rho_c$ and $\rho_c$ 
	is the critical closure density of the universe, 
	and $h$ is the Hubble constant in units of 100 km/sec/Mpc. 
	It is generally believed that most of the CDM 
	is made of weakly-interacting massive particles (WIMPs)
\cite{kt90}.  
	A commonly considered candidate for the WIMP 
	is the lightest neutralino, provided  
	it is the lightest supersymmetric particle (LSP) 
\cite{jkg}
	in the Minimal Supersymmetric extension of Standard Model (MSSM).  
	In most approaches the LSP is stable due to R-parity conservation
\cite{susyreview}. 
	The neutralino, being massive, neutral and stable, 
	often provides a sizeable contribution to the relic density,
	which is
	strongly model-dependent and varies by several orders of magnitude 
	over the whole allowed parameter space of the MSSM.
	The neutralino 
	relic density then can impose stringent constraints on the
	parameters of the MSSM, and may have 
	important consequences both for studies of SUSY 
	at colliders and in astroparticle experiments. 
	In light of this and taking into account the 
	continuing improvements in determining the abundance of CDM, 
	and other components of the Universe, which have now reached an
	unprecedented precision 
\cite{cmb},  
	one needs to be able to perform an accurate enough
	computation of the WIMP relic abundance, which would allow for a
	reliable comparison between theory and observation.
	On this way big progress in  
	calculations of the relic density 
	of neutralino in variety of supersymmetric models 
	has been already achieved 
[6--41].

	In the early universe neutralinos existed in 
	thermal equilibrium with the cosmic thermal plasma.
	As the universe expanded and cooled, the thermal energy is 
	no longer sufficient to produce neutralinos at an appreciable rate, 
	they decouple and their number density 
	scales with co-moving volume. 
        The sparticles significantly heavier than the LSP
	decouple at the earlier time and decay into LSPs. 
	Nevertheless there may exist some other 
	next-to-lightest sparticles (NLSPs) 
	which are not much heavier than the stable LSP. 
	The number densities of
	the NLSPs have only slight Boltzmann suppressions with respect to the
	LSP number density when the LSP freezes out of chemical equilibrium
	with the thermal bath.
	Therefore they may still be present in the thermal plasma, 
	and NLSP-LSP and NLSP-NLSP interactions hold LSP in thermal
	equilibrium resulting with significant reduction of the LSP
	number density.
	These {\em coannihilation}\ processes can be particularly 
	important when the LSP-LSP annihilation rate itself is suppressed
\cite{GriestSeckel,Mizuta:1993qp,EdsjoGondolo}.
	In the coannihilation with the LSP can be involved any SUSY particle,
	provided its mass is almost degenerate with the mass of the LSP
\cite{GriestSeckel,Belanger:2001fz}.
	In the low-energy effective MSSM (effMSSM), 
	where one ignores restriction from unification assumptions and 
	investigates the MSSM parameter space at the weak scale
\cite{EdsjoGondolo,Bottino,BKKmodel,Kim:2002cy}
	there is, in principle, no preference for the 
	next-to-lightest SUSY particle.

	The relativistic thermal averaging formalism 
\cite{GondoloGelmini} was extended to include coannihilation processes in 
\cite{EdsjoGondolo}, and was implemented in the DarkSusy code
\cite{Darksusy} for coannihilation of charginos and heavier neutralinos.
	In was found  
\cite{EdsjoGondolo} that 
	such a coannihilation significantly decreases the relic density. 
  	The importance of the neutralino coannihilation 
	with sferminos was emphasized and investigated for sleptons
\cite{EFOS-stau,Gomez:2000sj}, stops  
\cite{Boehm:2000bj,Ellis:2001nx} and sbottoms
\cite{Arnowitt:2001yh} in the so-called constrained MSSM (cMSSM) 
\cite{leszek,an93} or in supergravity (mSUGRA) models 
\cite{sugra}.

	In 
\cite{Bednyakov:2002js} 
	the comparative study of NCC and SLC channels, 
	exploration of relevant changes in the relic density 
 	and investigation of their consequences for detection of  
	CDM particles were performed in the effMSSM.
	This paper extends our investigations 
\cite{Bednyakov:2002js} to the 
	neutralino-stop and neutralino-sbottom coannihilations 
	and completes our consideration of the subject. 
	Therefore the calculations of neutralino relic density
	with inclusion of the all relevant coannihilation channels 
	(NCC, SLC, SQC) can be performed in the low-energy effMSSM.

\section{The \lowercase{eff}MSSM approach}
	As free parameters in the effMSSM, we use 
	the gaugino mass parameters $M_1, M_2$
	the entries to the squark and slepton mixing matrices 
	$m^2_{\tilde Q}, m^2_{\tilde U}, m^2_{\tilde D}, 
	 m^2_{\tilde R}, m^2_{\tilde L}$ 
	for the 1st and 2nd generations and 
	$m^2_{\tilde Q_3}, m^2_{\tilde T}, m^2_{\tilde B}, 
	 m^2_{\tilde R_3}, m^2_{\tilde L_3}$
	for the 3rd generation, respectively; 
	the 3rd generation trilinear soft couplings $A_t , A_b , A_\tau$; 
	the mass $m_A$ of the pseudoscalar Higgs boson, 
	the Higgs superpotential parameter $\mu$, and $\tan\beta$.
	To reasonably reduce the parameter space we assumed 
$ m^2_{\tilde U} = m^2_{\tilde D} = m^2_{\tilde Q}$,
$ m^2_{\tilde T} = m^2_{\tilde B} = m^2_{\tilde Q_3}$,
$ m^2_{\tilde R} = m^2_{\tilde L}$,
$ m^2_{\tilde R_3} = m^2_{\tilde L_3}$
	and have fixed $A_b = A_{\tilde \tau} = 0$
\cite{BKKmodel}. 
	The third gaugino mass parameter $M_3$ defines the 
	mass of the gluino in the model and is 
	determined by means of the GUT assumption $M_2 = 0.3\, M_3$.
	The remaining parameters defined our effMSSM 
	parameter space and were scanned randomly within the
	following intervals: 
\begin{eqnarray*} 
- 1 \tev < \!M_1\! < 1\tev, \ \  
-2\tev < \!M_2 , \mu , A_t\! < 2\tev, \ \  
1.5<\tan\beta < 50, \\  
50\gev < M_A < 1\tev, \quad
10 \gev^2 < m^2_{\tilde Q}, m^2_{\tilde L}, 
m^2_{\tilde Q_3}, m^2_{\tilde L_3} < 10^6\gev^2.
\end{eqnarray*}
	We have included the current experimental 
	upper limits on sparticle masses
	as given by the Particle Data Group 
\cite{pdg}. 
	The limits on the rare $b\rightarrow s \gamma$ decay 
\cite{flimits} have also been imposed. 
	The calculations of the neutralino-nucleon cross sections, and 
	direct detection rates follow the description given in 
\cite{jkg,BKKmodel}.
	The number density is governed by the Boltzmann equation
\cite{GondoloGelmini,EdsjoGondolo} 
\begin{equation} \label{boltzmann} 
{d n \over dt}+ 3 H n = - {\langle\sigma v\rangle} (n^2 - n_{\rm eq}^2 ) 
\end{equation} 
      	with $n$ either being the LSP number density if there are no other 
      	coannihilating sparticles, or 
	the sum over the number densities of all coannihilation partners.
	The index ``eq'' denotes the corresponding equilibrium value.
	To solve the Boltzmann equation 
(\ref{boltzmann}) one needs
	to evaluate the thermally averaged neutralino 
	annihilation cross section ${\langle \sigma v \rangle}$. 
	Without coannihilation processes
	${\langle \sigma v \rangle}$ 
	is given as the thermal average of the LSP 
	annihilation cross-section 
	$\sigma_{\chi\chi}$ times relative velocity $v$ 
	of the annihilating LSPs
${\langle \sigma v \rangle} = {\langle \sigma_{\chi\chi} v \rangle},$ 
	otherwise, it is determined as
	${\langle \sigma v \rangle} = 
	{\langle \sigma_{\rm eff} v \rangle}$, where  
	the effective thermally averaged cross-section
	is obtained by summation over coannihilating particles
\cite{GondoloGelmini,EdsjoGondolo}
\begin{equation} 
{\langle \sigma_{\rm eff} v \rangle} = 
\sum_{ij} {\langle \sigma_{ij}v_{ij} \rangle}
	 {n^{\rm eq}_i \over n^{\rm eq} }{n^{\rm eq}_j 
\over n^{\rm eq}}.
\end{equation}
	The relic density  is given by
$ 
\Omega = {m_\chi n_0} / \rho_{c}
$, 	
	where $n_0$ denotes the nowadays number density of the relics.
	For each point in the MSSM parameter space (MSSM model) 
	we have evaluated the relic density of the LSP
	$\Omega h^2$ \ under the following assumptions:
	ignoring any possibility of coannihilation (IGC), 
	taking into account only 
	neutralino-chargino (NCC), slepton (SLC), or squark (SQC) 
	coannihilations separately, 
	as well as including all of the 
	coannihilation channels (ACC).
	To this end in our former code 
\cite{BKKmodel} DarkSusy procedures of ${\langle \sigma_{\rm eff} v \rangle}$
	evaluation and solution of Boltzmann equation were implemented.
\begin{figure}[hb] 
\begin{picture}(100,105)
\put( 5,-64){\includegraphics{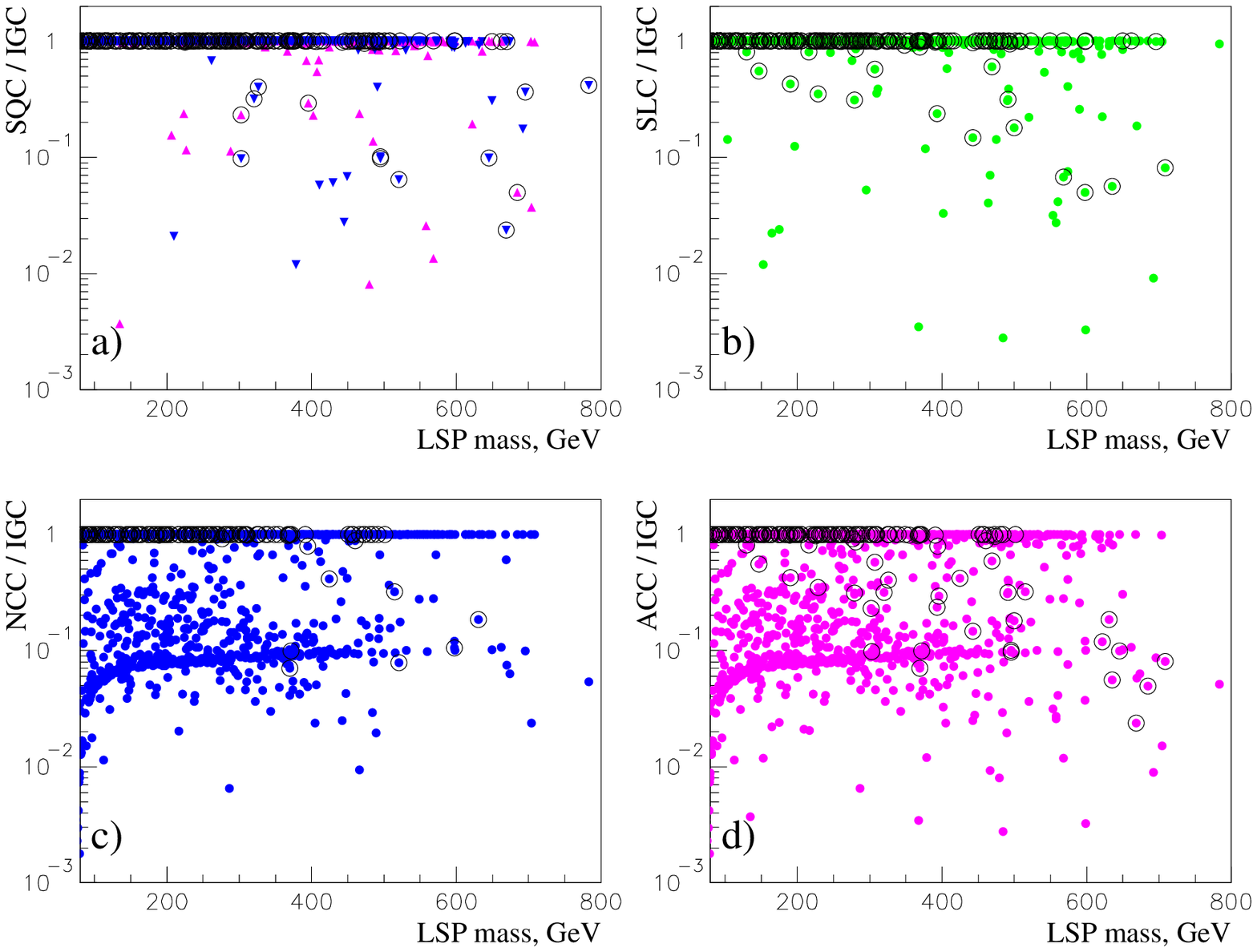}}
\end{picture}
\caption{Effects of  squark-neutralino (SQC), 
	            slepton-neutralino (SLC), and 
	neutralino-chargino\-(neutralino) (NCC) coannihilations in effMSSM. 
	Panels a)--d) display ratios 
	$\SQC / \IGC$, 
	$\SLC / \IGC$, 
	$\NCC / \IGC$, and 
	$\ACC / \SLC$ for the case when all coannihilations are included.
	The maximal reduction factors for all channels (NCC, SQC, and SLC) 
	are of the order of $10^{-3}$.
	Points in circles mark cosmologically interesting
	relic density $0.1< \COA < 0.3$. 
	In panel a) up-going triangles correspond to stop coannihilations
	and down-going triangles correspond to sbottom coannihilations. 
	One can see that stop and sbottom equally contribute. 
\label{A2I-f01}}
\end{figure} 

	Coannihilations with two-body 
	final states that can occur between neutralinos, charginos,
	sleptons, stops and sbottoms,
	as long as their masses are $m_i<2\mchi$ were included.
	The Feynman amplitudes for NCC, SLC and stop coannihilations
	were taken from DarkSusy 
\cite{Darksusy}, 
\cite{EFOS-stau}, and  
\cite{Ellis:2001nx}, respectively. 
	The amplitudes for the sbottom coannihilation  	
	were obtained on the basis of relevant stop amplitudes from  
\cite{Ellis:2001nx}.
	As in 
\cite{Bednyakov:2002js}
	the ${\langle \sigma_{\rm eff} v \rangle}$ 
	and ${\Omega h^2 }$ were calculated following 
	the relevant DarkSusy routines 
\cite{Darksusy}, which were added with with codes  
\cite{EFOS-stau}, and  
\cite{Ellis:2001nx}
	in a way that guarantees the correct inclusion of 
	SLC and SQC.
	We assume $0.1< \Omega h^2  < 0.3$ 
	for the cosmologically interesting region
\cite{acc}.

\section{Coannihilation effects in the relic density}
	The general view of the reduction effect on the relic density (RD)
	due to SQC, SLC, NCC and ACC are shown in 
Fig.~\ref{A2I-f01} as ratios $\COA / \IGC$.
	Here $\COA$ is 
	a common notation for $\ACC$,  $\NCC$, $\SQC$ or $\SLC$. 
	On the basis of our sampling (20000 models tested)
	the maximum RD suppression factor for NCC and SLC 
	channels is of the same order of about $10^{-3}$.
	Almost the same maximal suppression is also 
	for squark coannihilation channels.  
	These results depend on the 
	applied experimental limits on the 
	second-lightest neutralino, chargino and slepton
	stop and sbottom masses.
	The current experimental limits for 
	$m_{\tilde \tau}$, $m_{\tilde\mu}$, 
	$m_{\tilde\chi^\pm}$, 
	$m_{\tilde t}$, and
	$m_{\tilde b}$ are 80--90$\gev$ 
\cite{pdg},
	and therefore the critical LSP mass that enables 
	non-negligible NCC, SLC, and SQC contributions 
	is also of the same order ($\mchi \ge 80\gev$).
	From panel a) of the figure 
	one can conclude that stop (up-going triangles) 
	and sbottom (down-going triangles) equally contribute
	to reduction of RD due to coannihilations. 	
\begin{figure}[ht] 
\begin{picture}(100,80)
\put(18,-62){\includegraphics{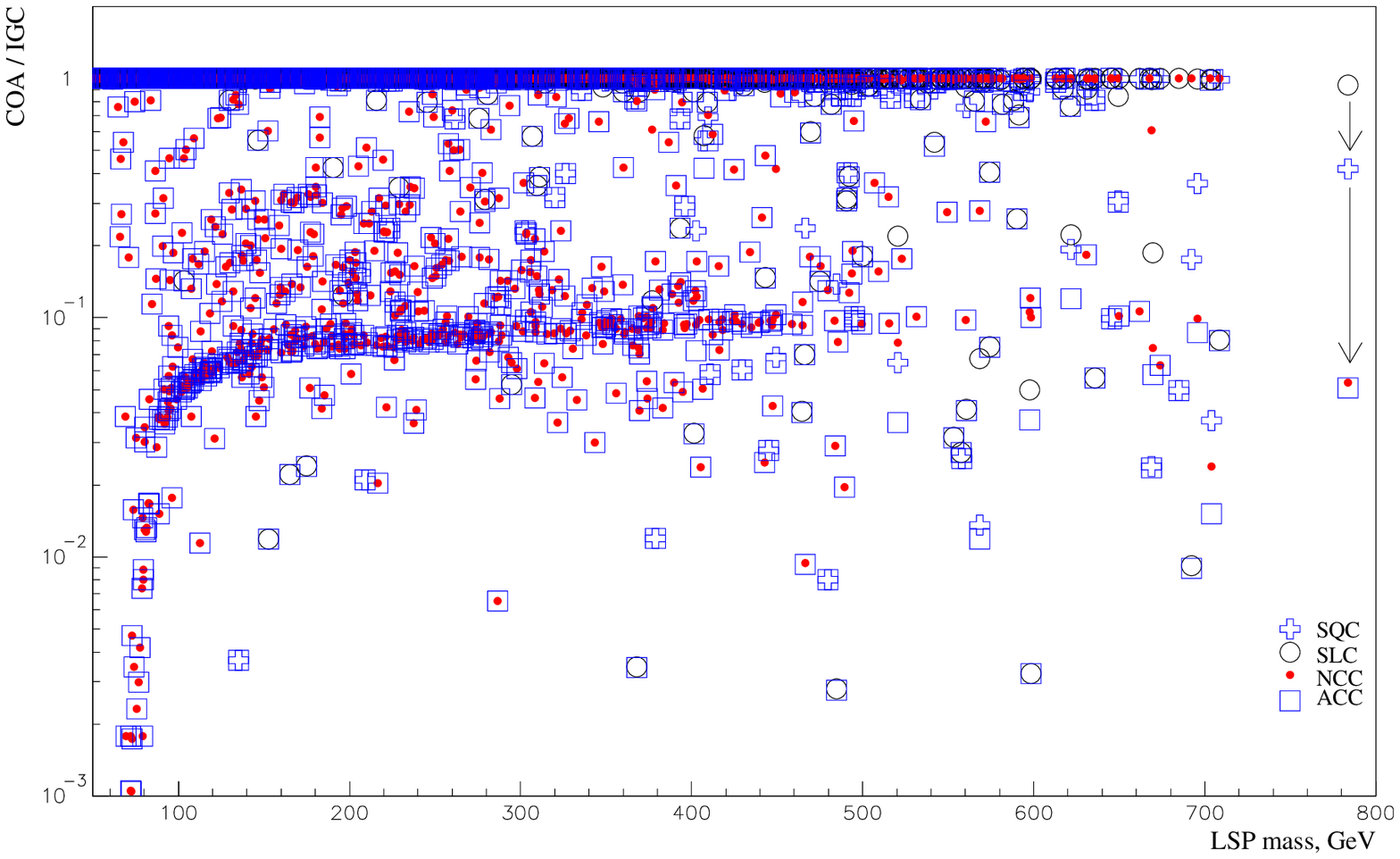}}
\end{picture}
\caption{The same as in Fig.~\ref{A2I-f01}, but plotted together.
 	Here 
	$\SQC / \IGC$, 
	$\SLC / \IGC$, 
	$\NCC / \IGC$, and 
	$\ACC / \SLC$  are marked 
	with crosses, circles, dots and squares, respectively. 
	Therefore, a square filled with a cross, circle, or dot 
	depicts a model that is affected only by SQC, SLC, or NCC, 
	respectively, and 
	any other coannihilation channel 
	gives negligible contribution.
	Such a situation takes place for the majority of models, 
	but there are some (quite few) models, given by empty squares, 
	for which at least two coannihilation channels are relevant. 
	For example, arrows in the right side of the figure 
	demonstrate how reduction of RD proceeds:
	SLC gives no effect ($\SLC / \IGC=1$),
	SQC reduces RD with factor $\SQC / \IGC\approx 0.4$,
	and finally NCC gives main contribution to RD suppression
	$\ACC / \SLC \approx\NCC / \IGC \approx 0.04$ 
	(the square nearly coincides with the dot).   
\label{A2I-f02}}
\end{figure} 

	The circles with symbols inside depict a some kind of ``constructive'' 
	reduction, when due to the coannihilations the relic density falls 
	into 
	cosmologically interesting region $0.1< \COA < 0.3$. 
  	Other points present the cases when coannihilations 
	too strongly reduce the relic density. 
	One can see that NCC plays the main role in
	``destructive'' reduction of RD,  
	these channels reduce maximal number of models 
	form cosmologically interesting region
\cite{EdsjoGondolo,Bednyakov:2002js}.
	Despite the fact, in ``constructive'' reduction 
	of RD all coannihilation channels 
	contribute at the same strength (there are almost the same 
	number of circled points in a)--c) panels).

\begin{figure}[ht] 
\begin{picture}(100,83)
\put(25,-46){\includegraphics{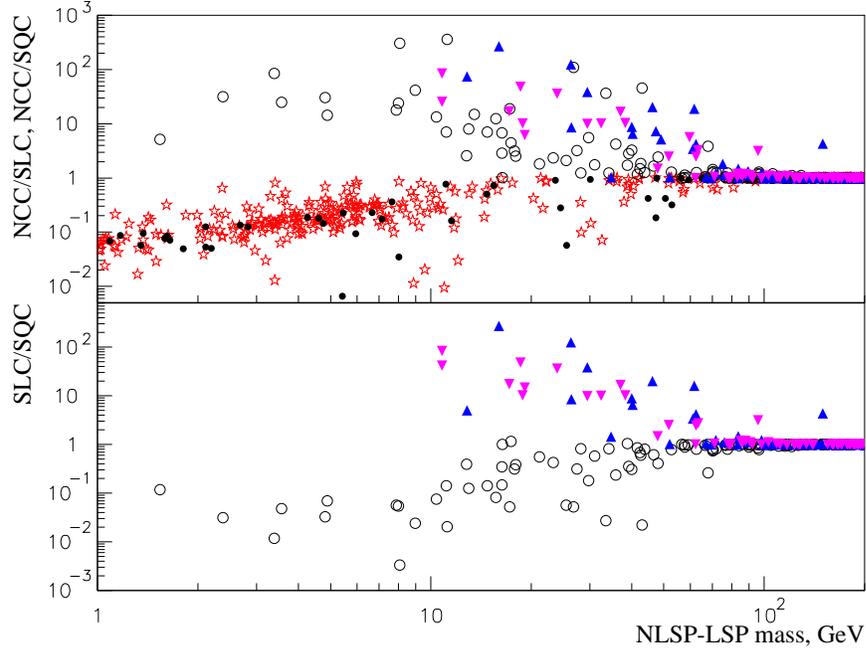}}
\end{picture}
\caption{Ratio $\NCC / \SLC$, $\NCC / \SQC$ and $\SLC / \SQC$ 
	versus $m^{}_{\rm NLSP} - \mchi$. 
	Open circle indicates that the $\tilde\tau$ is the NLSP, 
	star means that the light chargino $\tilde\chi^\pm$  
	is the NLSP, small filled square marks the model where the 
	second-lightest neutralino $\tilde\chi_2$ is the NLSP. 
	Up-going (down-going) triangles 
	indicates that $\tilde t$ ($\tilde b$) is the NLSP.
\label{MD-f04}}
\end{figure} 
	
	From 
Fig.~\ref{A2I-f02} one can see that the  reduction of RD 
	is mainly due to the only one dominant coannihilation channel
	NCC, SQC, or SLC. 
	The other channels of coannihilation in general play no role or 
	lead only to a much smaller further reduction
\cite{Bednyakov:2002js}.
	Figure
\ref{MD-f04} shows that for all coannihilation channels
	maximal RD reduction factors (less than 0.01)
	occur for mass differences 
	$m^{}_{\rm NLSP} - m^{}_{\rm LSP} \le 20\gev$.
	In contrast with NCC and SLC, 
	SQC can produce the same reduction effect with
	larger mass difference between squark and the LSP 
	($m^{}_{\tilde q} - m^{}_{\rm LSP}\approx 150\gev$)
	due to the possibility of coannihilation via 
	strong interactions.
	For NCC and SLC channels of coannihilation, 
	relevant effects occur if the mass difference between 
	the coannihilation partner and the LSP is within 15\%. 
	It was obtained that for SQC 
	the relevant effects occur if the mass difference between 
	the coannihilating squark and the LSP is within 50\%. 

  	Although other coannihilation processes 
	(including LSP annihilation with the next-to-NLSP (NNLSP) 
	and next-to-NNLSP, etc), 
	can in principal be also open from 	
Fig.~\ref{MD-f04} one can conclude that	
	dominant coannihilation channel is 
	defined by the type of the NLSP.	
	If next neutralino $\tilde\chi_2$ or 
	chargino $\tilde\chi^\pm$ is the NLSP, then  
	NCC indeed dominates. 
	Stau $\tilde\tau$ (or another slepton) being the 
	NLSP indeed entails a dominant SLC effect
\cite{Bednyakov:2002js}. 
	The SQC dominates when NLSP is the stop or the sbottom.

\begin{figure}[ht] 
\begin{picture}(100,105)
\put(13,-57){\includegraphics{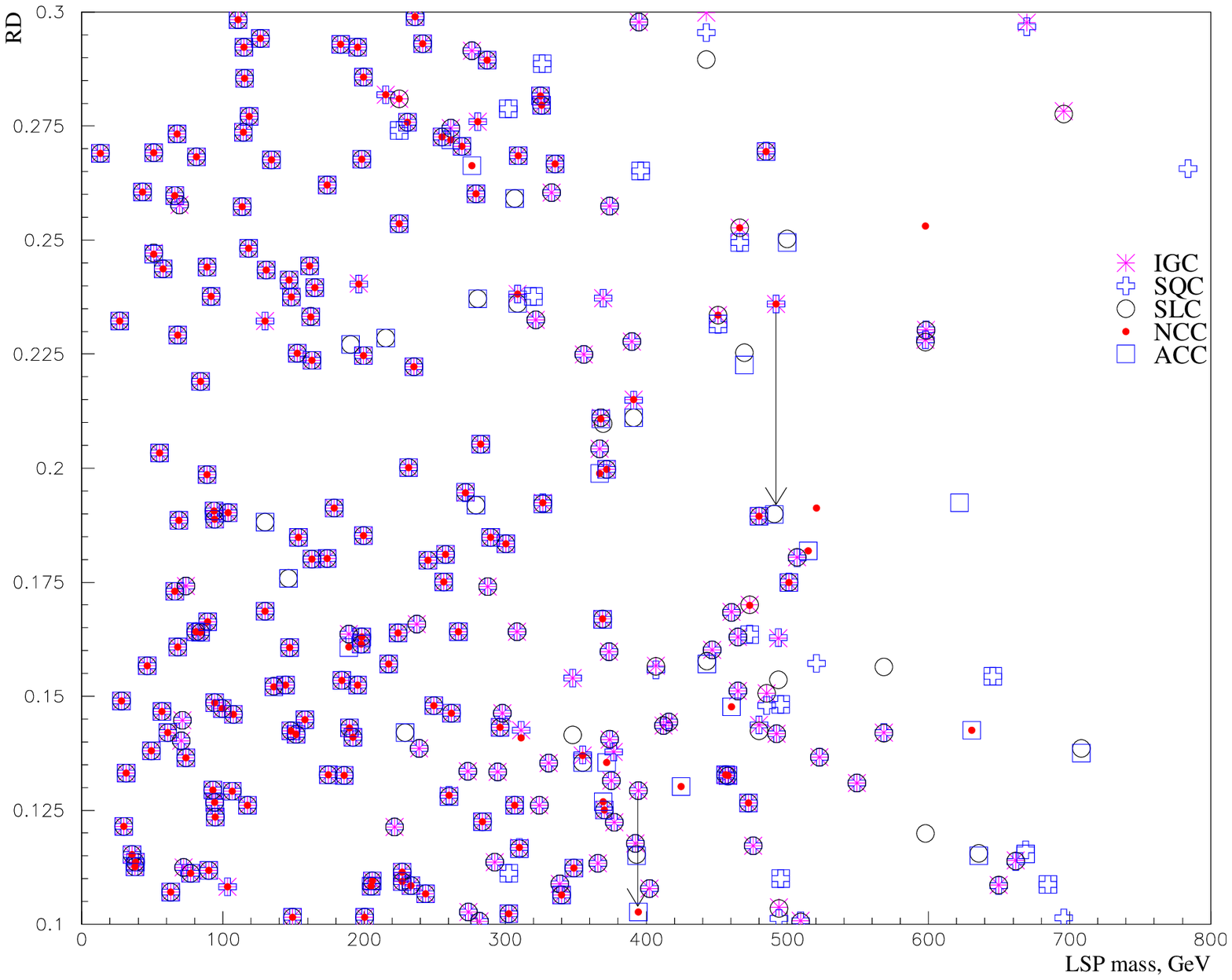}}
\end{picture}
\caption{Illustration of the shifting of effMSSM models inside 
	and outside the cosmologically interesting range $0.1< \COA < 0.3$ 
	due to NCC, SQC and SLC. RD 
	$\IGC$, $\SQC$, $\SLC$, $\NCC$ and $\ACC$ 
	are marked with stars, crosses, circles, small dots, and squares,
	respectively. 
	Therefore, a superposition of all symbols corresponds 
	to a model which is totally untouched by coannihilation. 
	A star-crossed circle 
	marks a model which 
	is untouched by SLC and SQC 
	($\SLC=\SQC=\IGC$), but shifted down due to NCC. 
	If the corresponding $\ACC$ (which is equal to $\NCC$) 
	remains within this range, 
	it still presents in the figure 
	below this star-crossed circle
	as an empty square with a black dot inside (see short arrow).
	By analogy, an square with a circle inside gives 
	a model which is shifted into the region due to SLC only
	($\ACC=\SLC$), and if the corresponding $\IGC=\NCC=\SQC$
	is also 
	in the cosmologically viable range, 
	it is located above the symbol 
	as an crossed star with a dot inside (see long arrow).
	A quite big amount of models is shifted out of 
	$0.1< \Omega h^2<0.3$ due to NCC (star-crossed circles).
\label{ACC-f05}}
\end{figure} 
	In Fig.
\ref{ACC-f05} all calculated 
	relic densities ($\IGC$, $\SQC$, $\SLC$, $\NCC$ and $\ACC$)
	are depicted in the cosmologically interesting region  
	$0.1< \COA < 0.3$.
	There is a quite big amount of models (mostly with $\mchi \le 250\gev$)
	which are completely unaffected by any kind of coannihilation.
	When at least one of coannihilation channels is relevant, 
	the RD decreases and some cosmologically unviable models with  
	$\IGC > 0.3$ enter the cosmologically interesting range
	$0.1< \COA < 0.3$,
	due to NCC (squares with a dot inside), 
	SLC (squares with circles inside),
	SQC (squares with crosses inside),
	or due to joint contribution of NCC, SQC, or/and SLC
	(empty squares). 
	There are also models which enter the less interesting  
	region for LSP to be CDM ($\COA < 0.1$).	
	The largest amount of models is shifted out 
	due to NCC (star-crossed circles),
	and a relatively small amount of models is shifted out
	due to SLC (crossed stars with a dot inside), 
	SQC (circles with a star and a dot inside), 
	both NCC and SLC (crossed stars).
	There are cosmologically interesting LSPs 
	within the full mass range $20\gev < \mchi < 720\gev$
(Fig.~\ref{ACC-f05}) accessible in our scan  whether or not
 	coannihilation channels are included.

\section{Coannihilation effects in the detection rates}
	Now we	consider the influence of all coannihilation
	channels in question (NCC, SQC and SLC) on prospects for  
	direct detection of CDM neutralinos. 
	We compare the rate predictions 
	for cosmologically interesting LSPs when the RD 
	is evaluated with or without any of coannihilation
	channel taken into account. 
\begin{figure}[ht] 
\begin{picture}(100,83)
\put(10,-70){\includegraphics{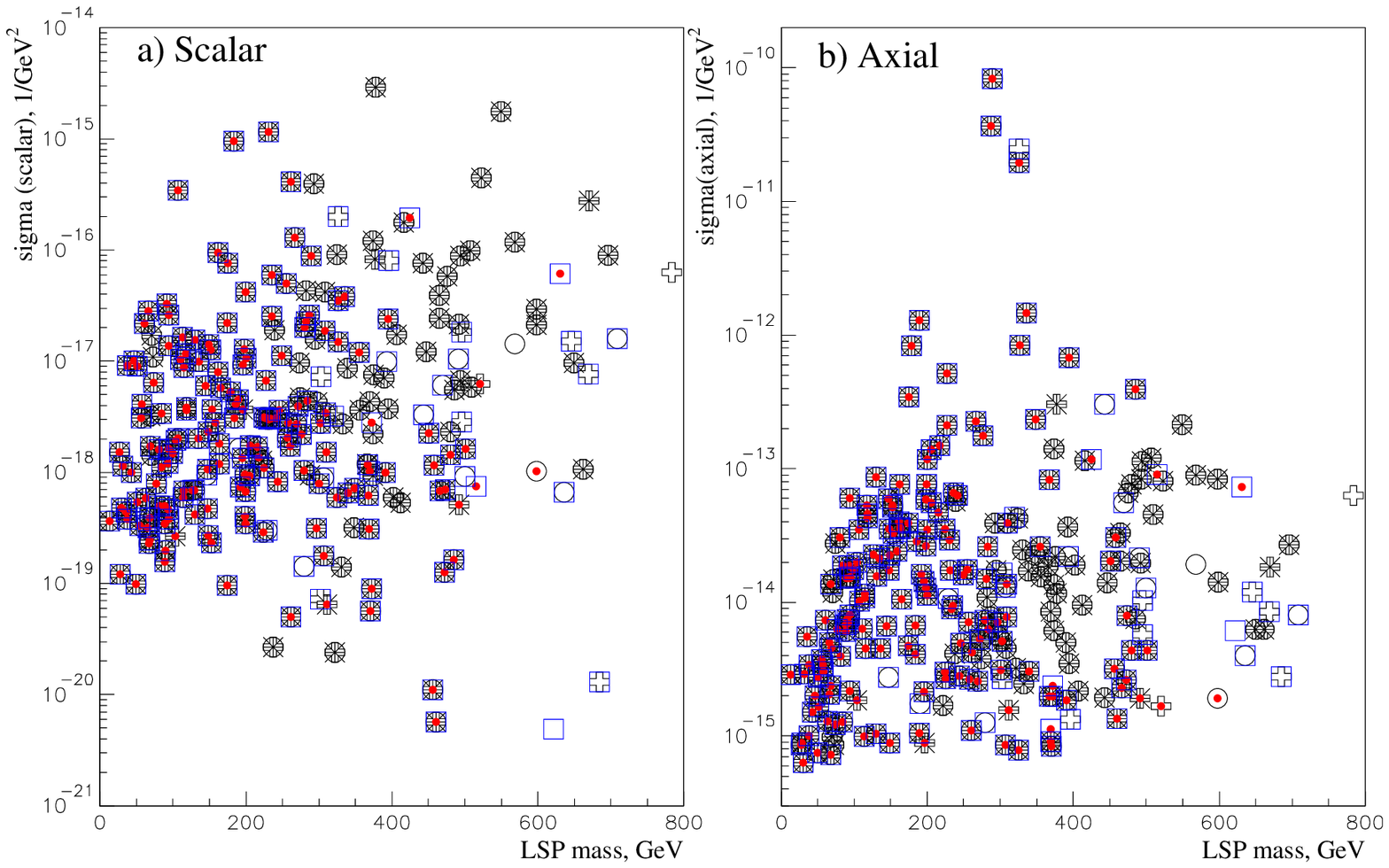}}
\end{picture}
\caption{Neutralino-proton scattering cross sections for 
	scalar (spin-independent) interaction
	(a) and axial (spin-dependent) interaction (b). 
	Stars, crosses, circles, small dots, and squares  correspond to
	$0.1<\IGC, \SQC, \SLC, \NCC, \ACC<0.3$, respectively. 
\label{CS-f08}}
\bigskip
\begin{picture}(100,70)
\put(13,-75){\includegraphics{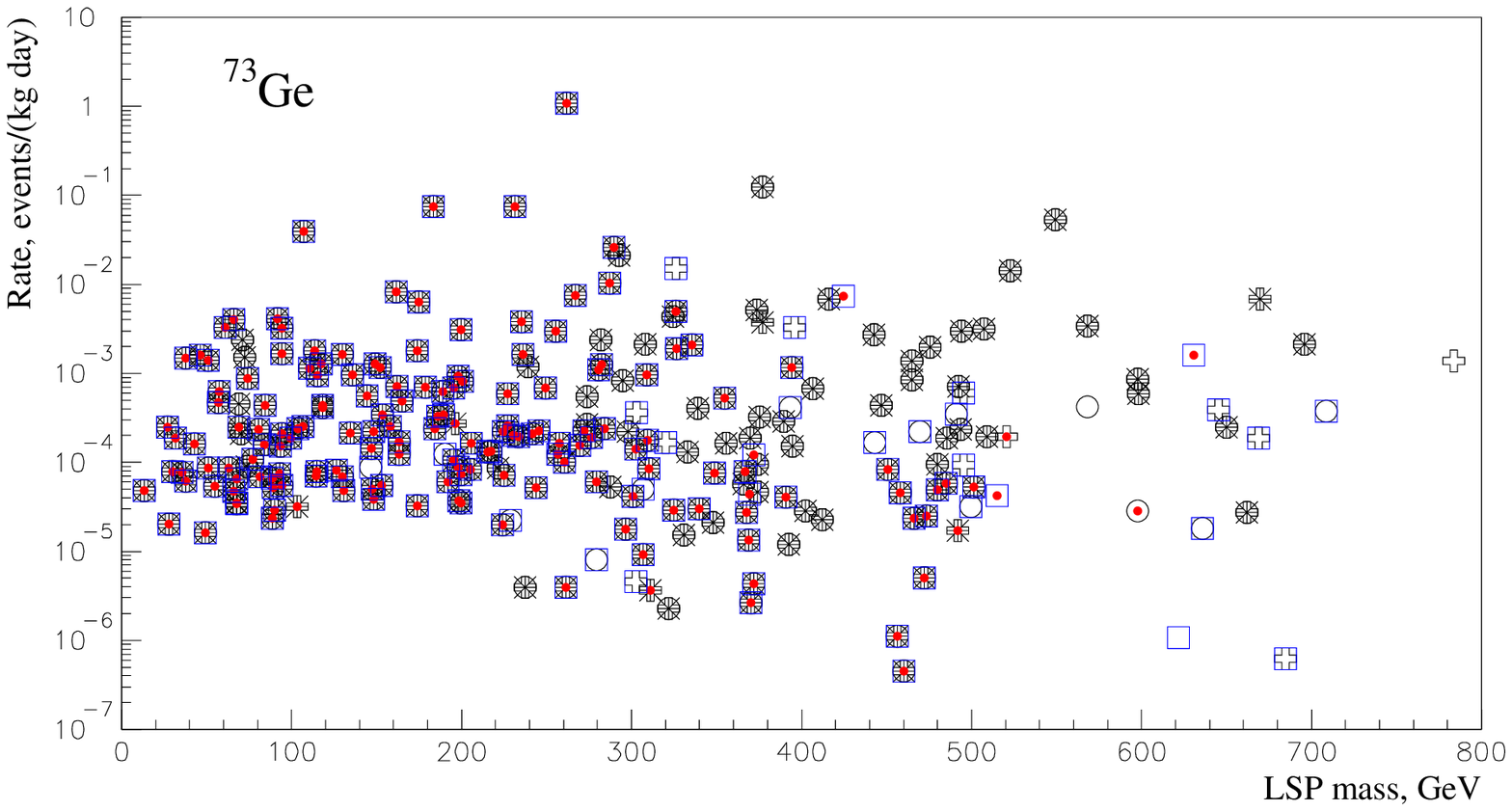}}
\end{picture}
\caption{Event rate for direct neutralino detection in a 
	$^{73}$Ge detector. 
	Stars, crosses, circles, small dots, and squares  correspond to
	$0.1<\IGC, \SQC, \SLC, \NCC, \ACC<0.3$, respectively. 
	NCC decreases the maximal event 
	rates for models with $\mchi \ge 400\gev$, 
	but the models with smaller LSP mass 
	are untouched by the coannihilations.
\label{RGe-f09}}
\end{figure} 
	We have seen
(Fig.~\ref{ACC-f05})	
	that the RD in most models with $\mchi \le 250\gev$ 
	is untouched by SQC, SLC and NCC, 
	mostly because the difference  $m_{\rm NLSP} - \mchi$ 
	is too large to yield significant effects, 
	therefore the corresponding detection rates are not influenced
	(depicted in the figures as square 
	filled with a star, a cross and a dot simultaneously). 
Figure~\ref{CS-f08} shows neutralino-proton 
	scattering cross sections for the scalar 
	(spin-independent) and the axial (spin-dependent) interactions.
	The models with $\mchi \le 250\gev$ are hardly 
	affected by coannihilation, and 
	for the majority of those models 
	both neutralino-proton and neutralino-neutron scattering 
	cross sections reach values  
	$\sigma_{\chi\,p }
	\le 10^{-17}\dtgev$ with the 
	maximal cross section of order $10^{-15}\dtgev$. 
	Cosmologically interesting models with 
	$\mchi \ge 250\gev$ were 
	influenced by coannihilations, 
	and the maximal value of the neutralino-nucleon cross-section 
	decreases from $10^{-15}\dtgev$ to $5\cdot 10^{-16}\dtgev$
	for the models with $\mchi > 500\gev$. 
	In total, 
	independently of neglection or inclusion of NCC, SQC and SLC
	the maximal scalar scattering neutralino-nucleon 
	cross section reaches 
	$10^{-16}$--$10^{-15}\dtgev$.	
	The spin-dependent neutralino-nucleon cross sections 
	are typically higher than the spin-independent ones, 
	and we have found the maximal values at 
$10^{-10}\dtgev$ for the axial neutralino-proton   and 
$10^{-11}\dtgev$ for the axial neutralino-neutron 
	scattering for the models 
	which are untouched by the coannihilations. 
	The majority of cosmologically interesting models yields 
	axial neutralino-proton scattering cross sections in the range 
	$5\cdot 10^{-16}\dtgev < \sigma_{\chi\,p}  < 2\cdot 10^{-12}\dtgev$ 
	and  
	axial neutralino-neutron scattering cross sections in the range 
	$2\cdot 10^{-16}\dtgev < \sigma_{\chi\,n}  <8\cdot 10^{-13}\dtgev $
\cite{Bednyakov:2002js}. 
	The SQC contribute in reduction of the cross sections, 
	but again not significantly. 
Figure~\ref{RGe-f09} shows the expected direct detection 
	event rates calculated for a $^{73}$Ge detector
	when NCC, SQC, SLC, and ACC are taken into account. 
	For models with $\mchi \le 250\gev$ 
	coannihilations of any kind play no role.
	The estimations of the event rate for models with 
	$\mchi \ge 400\gev$ are decreased mainly due to NCC
\cite{Bednyakov:2002js}.
\begin{figure}[h!] 
\begin{picture}(100,70)
\put(15,-75){\includegraphics{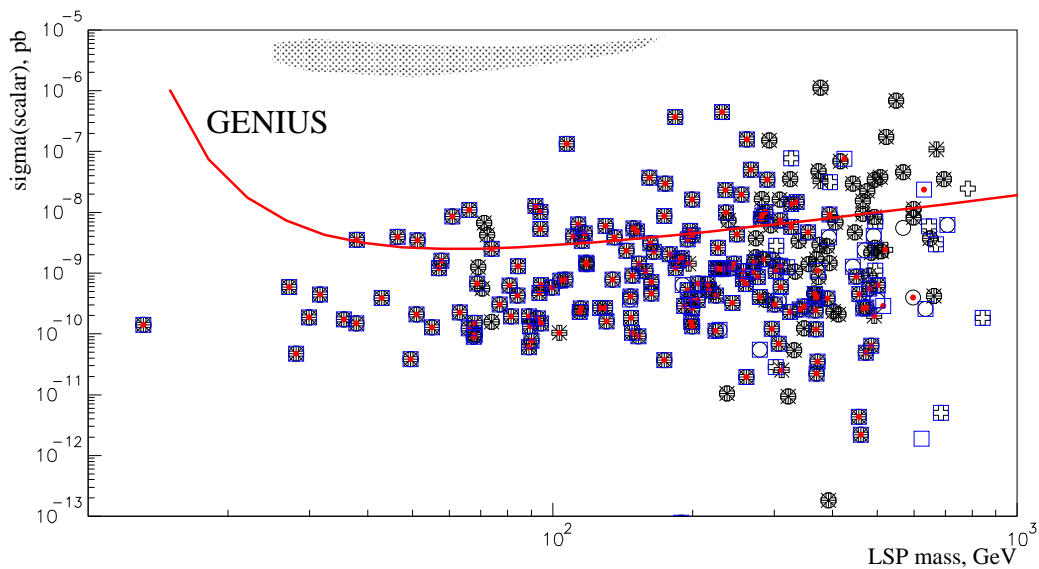}}
\end{picture}
\caption{Neutralino-proton scattering cross sections for 
	scalar (spin-independent) interaction.
	Expectations for GENIUS detector 
\protect\cite{GENIUS} and the 
	annual-modulation region of DAMA 
	(shaded region) 
\protect\cite{DAMA} 
	are also given.
	The maximal sensitivity of GENIUS
	and the region of DAMA are located 
	at $40 \le \mchi \le 300\gev$.
\label{GENIUS-f10}}
\end{figure} 
	
\section{Conclusion} 
	The neutralino relic density (RD) is calculated 
 	taking into account slepton-neutralino (SLC), 
	neutralino-chargino/neutralino (NCC), and 
	squark-neutralino (SQC)	coannihilation channels
	within the low-energy effective MSSM. 
	The maximum factors of RD decrease 
	due to NCC as well as due to SQC and SLC can reach $10^{-3}$, 
	as long as the lower experimental limits for 
	${m_{\tilde \tau}}$, ${m_{\tilde t}}$, ${m_{\tilde b}}$, 
	and $m_{\tilde\chi^\pm}$, 
	are of the order of $80\gev$.
	SQC, NCC, SLC produce comparable 
	RD reduction effects in the effMSSM.
	For the majority of models affected by coannihilations 
	it was observed that
	either NCC, SQC or SLC alone produces significant reduction of RD 
	while the other coannihilation channels can give  
	considerably smaller further reduction.
	Contrary to NCC and SLC, which 
	produce non-negligible effect only 
	if the NLSP mass is smaller than $1.15\mchi$, 
	for SQC the relevant NLSP mass could reach $1.50\mchi$. 
	The type of the NLSP determines the dominant coannihilation channel. 
	In the effMSSM 	all coannihilations do not imply 
	new cosmological limits on the mass of the LSP.
	The optimistic predictions for neutralino-nucleon 
	cross sections and LSP direct detection rates 
	for cosmologically interesting models 
	are almost untouched by these coannihilations. 
	Only for large $\mchi \ge 400\gev$, 
 	the respectively high values are reduced, 
	because of corresponding models are ruled out 
	from the cosmological interesting region $0.1<\IGC<0.3$.

	From 
Fig.~\ref{GENIUS-f10}
	one can see that the field of 
	maximal sensitivity of the best new-generation  
	CDM detectors, like GENIUS
\cite{GENIUS}, as well as the annual-modulation region of DAMA 
\cite{DAMA}  are located 
	at $40 \le \mchi \le 300\gev$, where coannihilation
	effects are almost invisible. 	
	Therefore 
	despite of obvious importance of sophisticated RD
	calculations including complete set of coannihilation
	channels it may happen that coannihilations will 
	play no any role at least for 
	{\em direct}\ detection of cold dark matter.

	This work was performed in collaboration with 
	prof. H.V.~Klapdor-Kleingrothaus and V.Gronewold.
	Author thanks Yudi Santoso for making his code available,
	I.V. Krivosheina for permanent interest to the work,
	the Max Planck Institut fuer Kernphysik for the hospitality 
	and RFBR 
	(Grants 00--02--17587 and 02--02--04009) for support.

\end{document}